# Current and perspective sensing methods for monkeypox virus: a reemerging zoonosis in its infancy

Ijaz Gul [a], Changyue Liu [a], Yuan Xi [a], Zhicheng Du [a], Shiyao Zhai [a], Zhengyang Lei [a], Chen Qun [a], Muhammad Akmal Raheem [a], Qian He [a], Zhang Haihui [a], Canyang Zhang [a], Runming Wang [a], Sanyang Han [a], Du Ke [b], Peiwu Qin [a,c*]

[a] Institute of Biopharmaceutical and Health Engineering, Tsinghua Shenzhen International Graduate School, Tsinghua University, Shenzhen, 518055, PR China

[b] Department of Mechanical Engineering, Rochester Institute of Technology, Rochester, NY, 14623, United States.

[c] Tsinghua- Berkeley Shenzhen Institute, Tsinghua Shenzhen International Graduate School, Tsinghua University, Shenzhen 518055, PR China

* **Corresponding author:** Peiwu Qin, Institute of Biopharmaceutical and Health Engineering, Tsinghua Shenzhen International Graduate School, Tsinghua University, Shenzhen 518055, PR China

**E-mail address**: pwqin@sz.tsinghua.edu.cn

# Abstract

## Objectives

The review is dedicated to evaluate the current monkeypox virus (MPXV) detection methods, discuss their pros and cons, and provide recommended solutions to the problems.

## Methods

The literature for this review is identified through searches in PubMed, Web of Science, Google Scholar, ResearchGate, and Science Direct advanced search for articles published in English without any start date until June, 2022, by use of the terms "monkeypox virus" or "poxvirus" along with "diagnosis"; "PCR"; "real-time PCR"; "LAMP"; "RPA"; "immunoassay"; "reemergence"; "biothreat"; "endemic", and "multi-country outbreak" and also, by tracking citations of the relevant papers. The most relevant articles are included in the review.

**Results**

Our literature review shows that PCR is the gold standard method for MPXV detection. In addition, loop-mediated isothermal amplification (LAMP) and recombinase polymerase amplification (RPA) have been reported as alternatives to PCR. Immunodiagnostics, whole particle detection, and image-based detection are the non-nucleic acid-based MPXV detection modalities.

**Conclusions**

PCR is easy to leverage and adapt for a quick response to an outbreak, but the PCR-based MPXV detection approaches may not be suitable for marginalized settings. Limited progress has been made towards innovations in MPXV diagnostics, providing room for the development of novel detection techniques for this virus.

## 1. Introduction

Monkeypox is a zoonotic disease caused by the monkeypox virus (MPXV) (Petersen et al., 2019). The MPXV is a large double-stranded DNA virus that belongs to the Orthopoxvirus genus (Sklenovská and Van Ranst, 2018). The MPXV was first reported in 1958 after twice pox-like disease outbreaks occurred in monkeys (Petersen et al., 2019). The rodents are considered a probable source of MPXV. The first case of MPXV in humans was reported in the Democratic Republic of Congo in 1970 (Kyaw et al., 2020). The virus has been predominantly reported in Central and Western Africa (Girometti et al., 2022). The modes of transmission include contact with body fluid, skin lesions, respiratory droplets, and fomite transmission (Kyaw et al., 2020). The approximate MPXV incubation period is about 5-21 days (Frey and Belshe, 2004). Recently, a multi-county monkeypox outbreak was reported to the World Health Organization (WHO) by several non-endemic countries. The number of cases is substantially increasing in Europe and many other countries. Since January 2022, and as of July 23, 2022, more than 16,000 cases and five deaths have been reported from 75 countries (WHO, 2022a). MPXV has now been declared a global health emergency by WHO (WHO, 2022a).

A number of methods have been reported for MPXV (Kulesh et al., 2004). Nucleic acid-based virus detection based on quantitative polymerase chain reaction (qPCR) and sequencing are widely used in clinical diagnosis and have also been used for MPXV detection (Meyer et al., 2004). Although sensitive and universal, the qPCR- and sequencing-based detection approaches are costly, which may hamper their practical implementation in marginalized settings. Isothermal amplification methods have been developed in order to overcome the drawbacks of PCR-based approaches (Glökler et al., 2021). Although isothermal amplification methods do not rely on thermal cyclers, each method has certain limitations in terms of selectivity and operational ease (Glökler et al., 2021), providing room for future developments. Since the MPXV is in the beginnings of spread, a review of MPXV detection techniques and possible development opportunities could be a timely addition to the fight against MPXV.

Herein, we highlight the MPXV detection modalities and discuss the challenges and opportunities. We start with a brief introduction to MPXV, including its genome organization, followed by a detailed discussion of monkeypox diagnostic approaches. The limitations and possible solutions are delineated.

## 2. Monkeypox Virus

Under an electron microscope, the MPXV (and other poxviruses) show a brick-shaped or oval geometry (Di Giulio and Eckburg, 2004). The size range of monkeypox virus is 200 to 250 nm (Alakunle et al., 2020). The whole virion is comprised of a core, lateral bodies, surface tubules, membrane, and outer envelope (**Figure 1 A**) (Frey and Belshe, 2004). The tightly packed core contains virus DNA (double stranded DNA), enzymes, and transcription factors.

The genome size of MPXV is about 197 kbp (Kugelman et al., 2014). The MPXV genome is mainly comprised of open reading frames (ORFs), hairpin loops, tandem repeats (TRs), and inverted tandem repeats (ITRs) (**Figure 1 B**) (Saxena et al., 2022). The genome is divided into a central genomic region and two terminal variable regions on both ends (Saxena et al., 2022).The central genomic region of MPXV and other orthopox viruses (OPVs) is homologous and conserved, involved in basic functions such as replication and assembly of virion, while the variable terminal regions are likely to be involved in virus virulence and immune evasion (Ramazan Azim et al., 2022; Shchelkunov et al., 2002).

The MPXV is genetically divided into two main clades; the West African (WA) clade and the Central African (CA) clade (Ramazan Azim et al., 2022). The fatality rate of the WA clade is 0-6% and a limited person-to-person transmission has been documented. The CA clade is more virulent (fatality rate is about 11 %) and is potentially more transmissible (Weaver and Isaacs, 2008). The WA clade has been reported in Nigeria, Ivory Coast, Sierra Leone, and Liberia, while the CA clade has been reported in the Republic of Congo, Democratic Republic of Congo, Cameroon, and Gabon. Recently, a new classification of the MPXV genome is proposed by (Christian et al., 2022). The MPXV genome is divided into clade 1, clade 2, and clade 3. The clade 1 corresponds to the genome from the CA clade, while clades 2 and 3 correspond to the prior WA clade (Christian et al., 2022). Another study also showed clustering of the MPXV genome into three clades: clade 1, 2, and 3 (Luna et al., 2022). The genomes of clade 3 are reported to belong to the last two outbreaks, *i.e*., 2017-19, and the recent 2022 outbreak. The clade 3 includes the human-to-human transmitting MPXV-1A clade (hMPXV-1A) and lineages including A.1, A.1.1, A.2, and B.1. The recent 2022 multi-country outbreak belongs to the B.1 lineage (Luna et al., 2022). The virus is increasingly spreading to non-endemic countries (**Figure 2**) (WHO, 2022b).

## 3. Monkeypox diagnosis approaches

### 3.1 Indirect Detection

The detection is based on virus-induced morphological changes to host cells or membranes.

#### i. Monkeypox diagnosis based on virus culture

Some viruses can induce macroscopic lesions (called pocks) on the chick chorioallantoic membrane (CAM). The pattern of pock formation, the time required for pock formation, and the size of the pock have been explored to differentiate different poxvirus infections, including MPXV (Magnus et al., 1959; Rondle and Sayeed, 1972). The morphological changes could be observed by a microscope or a naked eye. For instance, CAM was inoculated with MPXV, the pocks were visible and could be reckoned by the naked eye (Rondle and Sayeed, 1972). However, detection solely based on the above-mentioned characteristics may not be sufficient for accurate diagnosis.

The monkeypox isolates are grown in RK13 cells (Prier and Sauer, 1960), where cytopathic effects are observed within 24-48 h of infection. The major drawback of culture-based virus diagnosis is the prolonged assay time (Zhu et al., 2020), which is not suitable for mass testing scenarios. Further, virus culture methods need biosafety level 3 (BSL3) labs and pose a risk of laboratory-acquired infections (Artika and Ma'roef, 2017). Shell vial culture (SVC) has been developed as an alternative culture method for rapid *in vitro* detection of MPXV and other viruses (Zhu et al., 2020). In this method, the cell monolayer is grown on a cover slip in a shell vial culture tube and the specimen is inoculated on the monolayer, followed by low-speed centrifugation and immunofluorescence-based detection. The low-speed centrifugation step is introduced to enhance the virus's infectivity. The mechanical force resulting from low-speed centrifugation is thought to cause cell trauma, which subsequently enhances viral entry into cells, resulting in a reduced cell infection time (Jayakeerthi et al., 2006).

**ii. Diagnosis based on image analysis**

Image digitalization has already gained momentum for infectious disease diagnosis and monitoring (Pantanowitz et al., 2021). Chatbots have been developed for disease diagnostic evaluation and recommendation of immediate measures in case a patient contracts SARS-CoV-2 (Battineni et al., 2020). A monkeypox image dataset comprising of 43 original images and 587 images obtained after data augmentation is constructed (Ahsan et al., 2022b) (**Figure 3)**. Using the newly developed "Monkeypox 2022" dataset, an image classification model is proposed (Ahsan et al., 2022a). The study paves the way towards the development of image analysis-based tools for monkeypox virus detection.

## 3.2 Direct Detection

### i. Monkeypox immunodiagnostics

Hemagglutination test is a simple and cost-effective approach for virus detection. The test is based on the agglutination of erythrocytes in the presence of virus (Zhu et al., 2020). The hemagglutination mechanism led to the development of another assay called the hemagglutination inhibition (HI) assay (Hierholzer et al., 1969). The HI approach relies on virus-specific antibodies to detect viral antigens. The MPXV strains are tested using hemagglutination and HI tests (Rondle and Sayeed, 1972). The test cannot differentiate MPXV from variola and vaccinia viruses, but can differentiate cowpox from MPXV. The test can be used to estimate the evolutionary relationship of viral strains or species.

The enzyme linked immunosorbent assay is a widely used protein detection method (Sadeghi et al., 2021). A commercially available Orthopox BioThreat® Alert Assay for orthopox virus (OPV) detection is a reliable OPV detection method (Townsend et al., 2013). This antibody-based lateral flow assay captures virus antigens and detects viral load at $10^4$ PFU/mL. The surface protein A27 is found to be the most immunogenic protein for virus particle capture and detection (Stern et al., 2016b). After a comprehensive screening of A27 binding antibodies, an ELISA approach is developed for orthopox viruses, including MPXV. The method's detection limit is $1 \times 10^3$ PFU/mL. In a similar line of work, an ABICAP (Antibody Immuno Column for Analytical Processes) immunofiltration system is developed by Stern *et al*. The system has an OPV detection sensitivity of $10^4$ PFU/mL with an assay time of 45 minutes (Stern et al., 2016a). A dot-immunoassay based on protein array technology can detect MPXV at a concentration range of $10^3$–$10^4$ PFU/mL within 39 minutes (Poltavchenko et al., 2020). Recently, Ulaeto *et al*. describe the characteristics of an LFA for detection of orthopoxviruses (Ulaeto et al., 2022). The assay detects vaccinia virus samples spiked in human saliva and clinical sample buffer with a detection limit of between $10^{4.5}$ to $10^5$ PFU /mL within 20 minutes. Since this assay detects orthopoxviruses, the test can be further explored for MPXV detection in real samples. Combining clinical presentation of MPXV with the LFA test could provide a rapid MPXV detection tool. The above-mentioned immunodetection modalities are suitable for generic orthopox virus detection applications, but none of them is selective for MPXV.

### ii. Whole particle detection

Finding biomarkers for a newly emerged virus is challenging and may hamper direct implementation of routine diagnostic methods. In this regard, whole particle detection using electron microscopy (EM) is a powerful alternative (Hughes et al., 2017). Transmission electron microscopy is a good first step for the detection of viruses as it provides information about the shape and amount of virus load with a small sample volume (Richert-Pöggeler et al., 2019). The use of virus-specific antibodies in immuno-electron microscopy (IEM) further improves the detection accuracy of EM (Lavazza et al., 2015).

The EM has been used to detect monkeypox and other orthopox viruses (Gelderblom and Madeley, 2018). Although EM is suitable for laboratory validation of the virus detection results, the approach has certain limitations, such as high cost of the instrument, the requirement of highly trained staff, and low sample throughput (Gelderblom and Madeley, 2018).

**iii. Detection by genome sequencing**

Genome sequencing is the gold standard to identify novel or mutated viruses. The genome sequencing not only identifies the target virus but may pinpoint the presence of other viruses in the sample that can help to plan a treatment plan for a particular disease. MPXV detection based on qPCR coupled with genome sequencing has been reported (Dumont et al., 2014). To date, 200 genome sequences of MPXV isolates from recent outbreak in non-endemic countries have been reported (NCBI, 2022). Whole genome sequencing is a time-consuming process and needs expensive instruments, trained staff, and skilled bioinformaticians for computational analyses. These limitations need to be overcome to harness the potential of genome sequencing approaches.

**iv. Monkeypox virus detection based on PCR**

The polymerase chain reaction (PCR) is widely regarded as the gold standard for nucleic acid detection. According to WHO recommendations, PCR (conventional or real-time) is a standard method for MPXV laboratory validation (WHO, 2022c). The detection can be combined with sequencing or other orthopox detection assays (Saxena et al., 2022).

Conventional PCR-based MPXV detection involves PCR amplification and restriction digestion of the PCR amplified fragments to identify MPXV based on restriction fragment length polymorphism.

A hemagglutinin PCR (HA-PCR) assay is developed based on MPXV specific primers coupled with *Taq*I restriction digestion (Ropp et al., 1995). The method could not distinguish different MPXV isolates. To improve the detection accuracy of the PCR assay, a A-type inclusion body protein (ATI)  gene has been used to detect MPXV and other orthopox viruses based on PCR-based gene amplification and *XbaI* digestion (Meyer et al., 2004, 1997). The method can differentiate MPXV strains based on restriction digestion. In another development, the open reading frame (ORF) of the ATI gene is identified, sequenced, and compared with other related poxviruses (Neubauer et al., 1998). The unique deletions are found in the OFR of MPXV that are harnessed for specific detection of the MPXV ATI gene. This PCR method differentiates 19 MPXV stains. The specificity is confirmed by *Bgl*II restriction digestion.

Compared to traditional PCR, real-time PCR is rapid and sensitive. Due to the low GC content and 90% similarity with other Eurasian *orthopoxviruses*, designing an MPXV-specific TaqMan assay is challenging. Li *et al*. developed a real-time PCR assay where minor groove binding protein-based (MGB) probes are developed (Li et al., 2006). The use of MGB stabilizes probe-template interaction, enables the use of small probe sequences for single nucleotide polymorphism (SNP) detection, and enhances assay sensitivity and specificity ( Belousov et al., 2004; Li et al., 2006). The method could detect 15 MPXV isolates at a 10 ng concentration. The assay efficiency with freshly diluted DNA is 97%, while it is reduced to 67% after multiple freeze-thaw cycles. These observations indicate that a fresh sample should be used in order to achieve maximum assay efficiency. Detection of MPXV and other orthopox viruses based on melting-curve analysis (MCA) has also been reported (Gelaye et al., 2017; Nitsche et al., 2004; Olson et al., 2004). Both clades (West African and Congo Basin) of MPXV have 99% sequence similarity (Simpson et al., 2020), but are significantly different in terms of virulence (Simpson et al., 2020), it is a big challenge to develop a clade-specific real-time PCR detection approach due to the limited availability of unique sequences. In an effort to differentiate between isolates from two different clades, the terminal genomic sequences of MPXV strains are analyzed (Li et al., 2010b). Since the terminal sequences show relatively more sequence variability than the central genomic region and the tumor necrosis factor (TNF) gene lies in the terminal genomic region, the TNF gene is chosen to design primers and probes for the West African MPXV specific assay called G2R-WA. No unique sequences are found in the TNF gene of the Congo Basin clade. Therefore, another gene (C3L) is targeted for Congo Basin MPXV (Li et al., 2010b).

Multiplex detection can significantly reduce the misidentification of co-existing pathogens (Hughes et al., 2021; Luciani et al., 2021a). A multicolor, multiplex approach for MPXV detection is reported where MPXV is specifically detected in the presence of variola virus (VARV) and VZV (Maksyutov et al., 2016). The target genes harboring unique sequences for MPXV, VARV, and VZV are F3L, B12R, and ORF38, respectively. The specificity of the developed approach is 100% and the LODs of 20 copies per reaction for MPXV and VARV, and 50 copies per reaction for VZV, are reported. The robustness of the approach is demonstrated by successfully detecting the different combinations of MPXV, VARV, and VZV samples.

The standard poxvirus detection approach combines the disease’s clinical symptoms with a poxvirus generic PCR assay, followed by a poxvirus-specific PCR assay (Olson et al., 2004). These pan-pox real-time PCR methods are instrumental in the accurate diagnosis of poxvirus infection. Based on GC content, the chordopoxviruses (poxviruses that infect vertebrates) of the subfamily *Chordopoxvirinae* have two distinct genome types: one genome type contains high GC content (> 60%), while the other genome type is comprised of low GC content (30-40%) (Li et al., 2010a). GC-content-based pan-pox PCR assays are developed by (Li et al., 2010a). The assays are termed high-GC PCR and low-GC PCR assays. The developed PCR assays detected DNA samples from more than 150 isolates and strains of chordopoxviruses. The detection approach is based on conventional PCR, and PCR amplicons are evaluated by *Taq*1 RFLP patterns. In a similar line of work, a real-time PCR assay for universal detection of orthopoxviruses is reported (Luciani et al., 2021b). The system is reported to be able to detect poxviruses excluded in a previous study (Li et al., 2010a) as well as those from the subfamily *Entomopoxvirinae*. This assay targets a 100bp highly conserved sequence in the D6R gene of poxviruses. The specificity of the assay for vertebrate samples is 99.8%, while it is 99.7% for arthropod samples. The system is 100% sensitive for vertebrate samples and 86.6% for arthropod samples. The detection limits are reported to be 100 or 1000 copies per reaction, depending on the poxvirus species.

**v. Detection based on isothermal amplification**

More than ten types of different isothermal amplification methods have been reported and demonstrated for nucleic detection (Glökler et al., 2021). Loop-mediated isothermal amplification (LAMP) and recombinase polymerase amplification (RPA) are well-explored isothermal nucleic acid amplification-based virus detection methods (Becherer et al., 2020; Huang et al., 2020). The LAMP technology relies on two internal primers called forward internal primer (FIP) and backward internal primer (BIP), two outer primers known as forward outer primer (F3) and backward outer primer (B3), and a DNA polymerase with strand displacement activity (Becherer et al., 2020). The reaction is carried out at 60-65 °C. The amplification reaction is accelerated by using two loop primers, the forward loop (LF) and the backward loop primer (LB) (Nagamine et al., 2002). The annealing of FIP which has two target sequences (separated by a spacer) complementary to the two different regions of the template, initiates strand synthesis and elongation (**Figure 4A**). Subsequently, the F3 primer displaces the FIP strand, producing a single-stranded DNA (ssDNA) strand that is used as a template by BIP (**Figure 4B**). The BIP, which also has two target sequences complementary to the template DNA at two different regions, starts strand elongation of the ssDNA template that is later displaced by B3 (**Figure 4C**). The 5' and 3' ends of the template DNA have inward complementary sequences, forming a stem-looped DNA that is exponentially amplified by loop primers (**Figure 4C and D**). LAMP-based MPXV clade-specific assays are developed where West African (the assay named W-LAMP) and Congo Basin MPXV (the assay named C-LAMP) clades are selectively detected (Iizuka et al., 2009). A turbidimeter is used to analyze the LAMP reaction, and restriction digestion is used to confirm the LAMP products. Although promising, the assay needs a 60-minutes reaction time and 6 primers. Furthermore, primer design is relatively complex. To overcome these limitations, RPA has been proposed as an attractive alternative (Lobato and O'Sullivan, 2018) (**Figure 5)**. RPA signal is detected by gel electrophoresis, real-time monitoring (Fan et al., 2020) or by lateral flow assay (Fan et al., 2020). In the case of real-time detection, the fluorogenic probe along with primers is added to the reaction system where cleavage of the probe by exonuclease leads to a fluorescent signal. The RPA-based MPXV detection shows satisfactory results with reduced assay time and reagent cost (Davi et al., 2019). Summary of MPXV detection methods (**Table 1)**.

### vi. WHO sample collection guidelines

According to WHO's guidelines (WHO, 2022d), the specimen type can be: (a) skin lesion material including swabs of lesion exudate, lesion roofs, and lesion crusts; (b) oropharyngeal swabs; (c) rectal and or genital swabs; (d) urine; (d) semen; (d) whole blood; (d) serum, and (f) plasma. Skin lesion material and oropharyngeal swabs (if feasible) are recommended for diagnosis purposes, while serum and plasma samples can be obtained for aid in diagnosis or for research purposes. The rest of the specimen types are recommended to be collected for research purposes that are subject to ethics guidelines. The sample can be refrigerated (for 7 days) or frozen at -20 °C or below (for 60 days).

## 4. Conclusions and Prospects

The reemergence of MPXV after COVID-19 is a clear indication that timely detection of viruses is instrumental in controlling the onset and spread of outbreaks. PCR is a gold standard for nucleic acid detection. Although sensitive and selective, the PCR-based MPXV detection approaches are not suitable for resource-constrained settings. Alternatively, isothermal nucleic acid amplification techniques are emerging alternatives. The development timeline of MPXV diagnostics indicates that limited progress has been made towards innovations in MPXV diagnostics, indicating an obvious research gap. The WHO recommends the development of point-of-care (POC) devices. Internet of things (IoT)-based POC devices have attracted substantial attention. The IoT-based LAMP for MPXV detection would be a promising approach. IoT-based detection of COVID-19 using LAMP technology has been demonstrated with satisfactory performance (Nguyen et al., 2022). Similarly, another field-deployable RT-LAMP-based device for onsite virus inactivation and detection is reported by (Ge et al., 2022). These advanced approaches can be extended to MPXV diagnostics. Although a number of nucleic acid methods based on LAMP technology have been developed, the approach needs a high temperature and six primers.

Alternatively, RPA technology can be equipped with smartphones for field applications since RPA needs two primers and the reaction can be performed at 37-42 °C. Although promising, the approach has some limitations. For instance, RPA, like PCR, can be inhibited by a high concentration of genomic DNA (Rohrman and Richards-Kortum, 2015). Furthermore, multiplex detection using RPA might be challenging as RPA primers for different genes or targets compete for the RPA proteins. The problems can be solved by integrating RPA with microfluidic platforms where multiplex detection could be performed in separate microfluidic compartments (Ahn et al., 2018; Li et al., 2019). The use of multiple quantum dots for different targets and coupling with DNA barcodes could be a fascinating approach to develop a POC detection system where MPXV could be distinguished from the rest of the poxviruses. Further, a separate solid-phase amplification can also overcome the problems of RPA-based multiplex detection (del Río et al., 2014).

Wearable devices have found increased application in recent years (Chaudhury et al., 2022). Since MPXV has well-defined signs and symptoms, the data obtained from smartphone apps and smartwatches can be used to predict the pre-symptomatic cases. For instance, Mishara *et al*. report a comprehensive study where physiological data from smartwatches is used to predict COVID-19 pre-symptomatic cases (Mishra et al., 2020). Inspired by this work, many machine learning algorithms have been reported in recent studies (Cho et al., 2022; Shapiro et al., 2021) and are equally important for MPXV detection. It is important to know the virus's infectiousness status after infection. The available methods solely predict the presence or absence of virus or virus particles. A method for determining the virus's infectiousness status in infected patients could be a valuable addition to MPXV research. For this, immunodiagnostic methods are promising but do have certain limitations, especially poor selectivity for MPXV. The antigen detection methods are rapid and cost-effective but less sensitive. Same implies to MPXV immunodiagnostics. Therefore, novel MPXV antigen detection methods are being developed in the near future. Due to MPXV's similarity with other orthopox viruses, finding a unique antigen is a daunting challenge. The E8L protein of MPXV is a membrane protein and is a potential target for vaccines. Recently, non-cross-reactive epitopes for MPXV are reported within the E8L protein via a computational approach (Gao and Gao, 2022). It is anticipated that E8L binding peptides can also be discovered in a similar way and used as a biosensing layer for specific detection of MPXV. Further, the E8L binding aptamers and nanobodies (Guo et al., 2021) can be valuable contribution. To the best of our knowledge, the MPXV entry receptor is still unknown; the discovery of the MPXV entry receptor and the development of MPXV sensors based on the entry receptor could be a useful future work. Novel MPXV biosensors can be developed based on photonics (Shin et al., 2015), quantum dots (Ming et al., 2015), electrochemiluminescence (Nikolaou et al., 2022), electrochemical transduction (Liao et al., 2022; Wu et al., 2022), lab on a chip(Zhu et al., 2020) , and CRISPR technology (Wu et al., 2022).

## Conflict of Interest

The authors declare that they have no known competing financial interests or personal relationships that could have appeared to influence the work reported in this paper.

## Funding Source

Authors gratefully acknowledge funding from National Natural Science Foundation of China 31970752, Science, Technology, Innovation Commission of Shenzhen Municipality JCYJ20190809180003689, JSGG20200225150707332, JSGG20191129110812, Shenzhen Bay Laboratory Open Funding SZBL2020090501004.
The sponsors have no role in the study design, analysis, writing of the manuscript, and in the decision to submit the manuscript for publication.

## Ethical Approval Statement

This work does not involve the use of human or animal subjects.

Figure 1

A

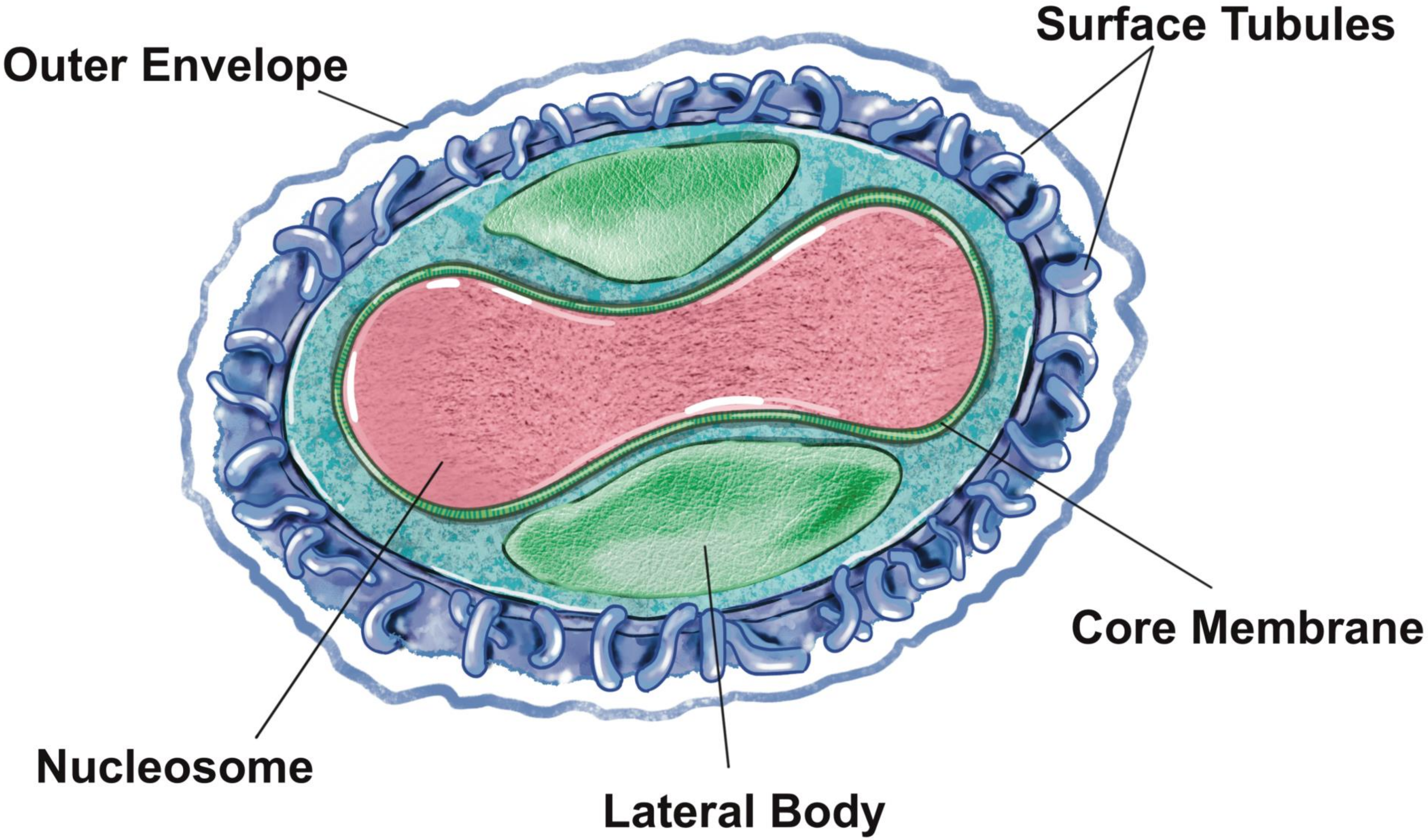
Outer Envelope
Surface Tubules
Core Membrane
Nucleosome
Lateral Body

B

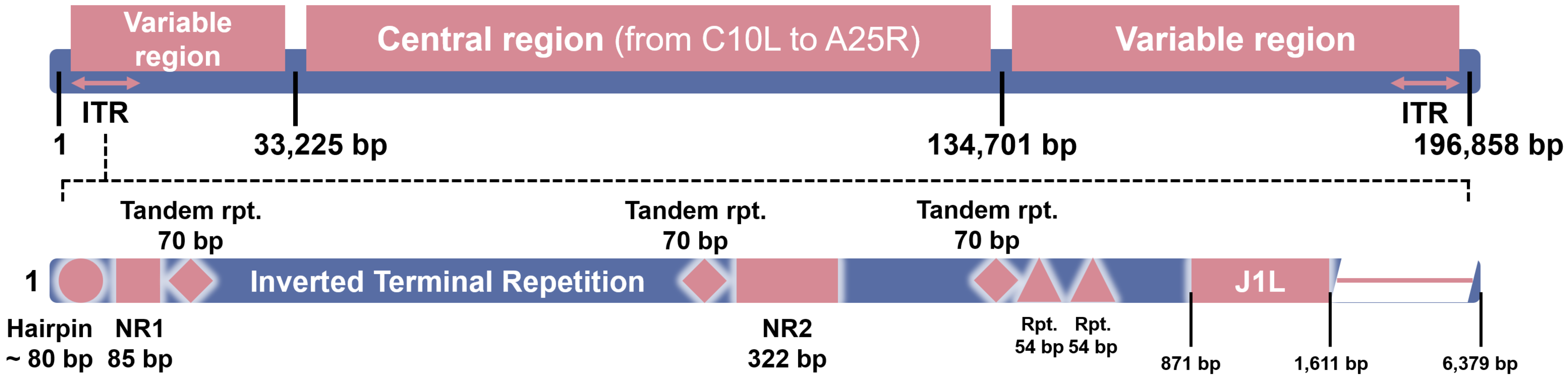

Figure 2

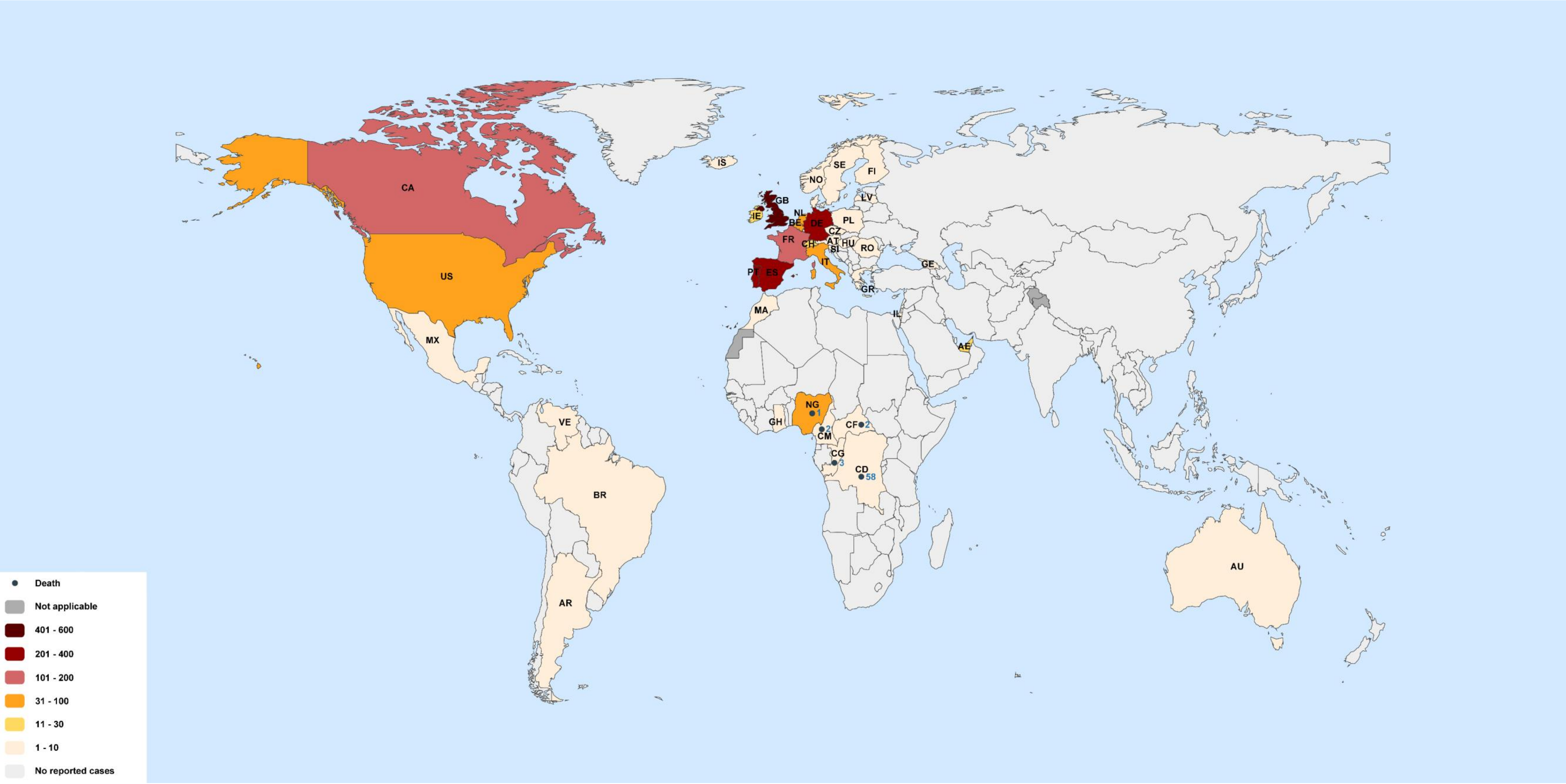

Figure 3

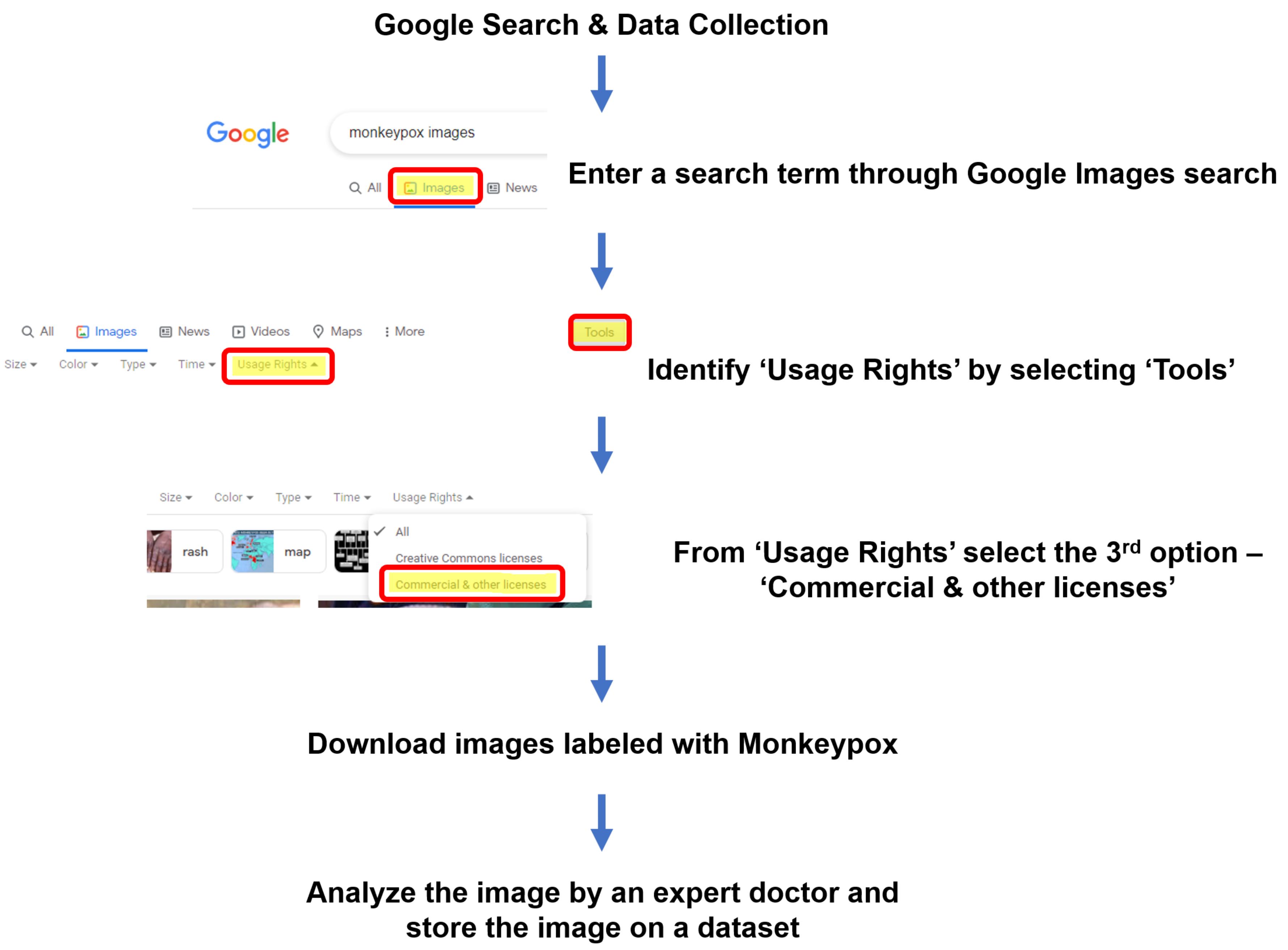

Google Search & Data Collection
Google
monkeypox images
All
Images
News
Enter a search term through Google Images search
All
Images
News
Videos
Maps
More
Tools
Size
Color
Type
Time
Usage Rights
Identify 'Usage Rights' by selecting 'Tools'
Size
Color
Type
Time
Usage Rights
All
Creative Commons licenses
rash
map
Commercial & other licenses
From 'Usage Rights' select the 3rd option – 'Commercial & other licenses'
Download images labeled with Monkeypox
Analyze the image by an expert doctor and store the image on a dataset

Figure 4

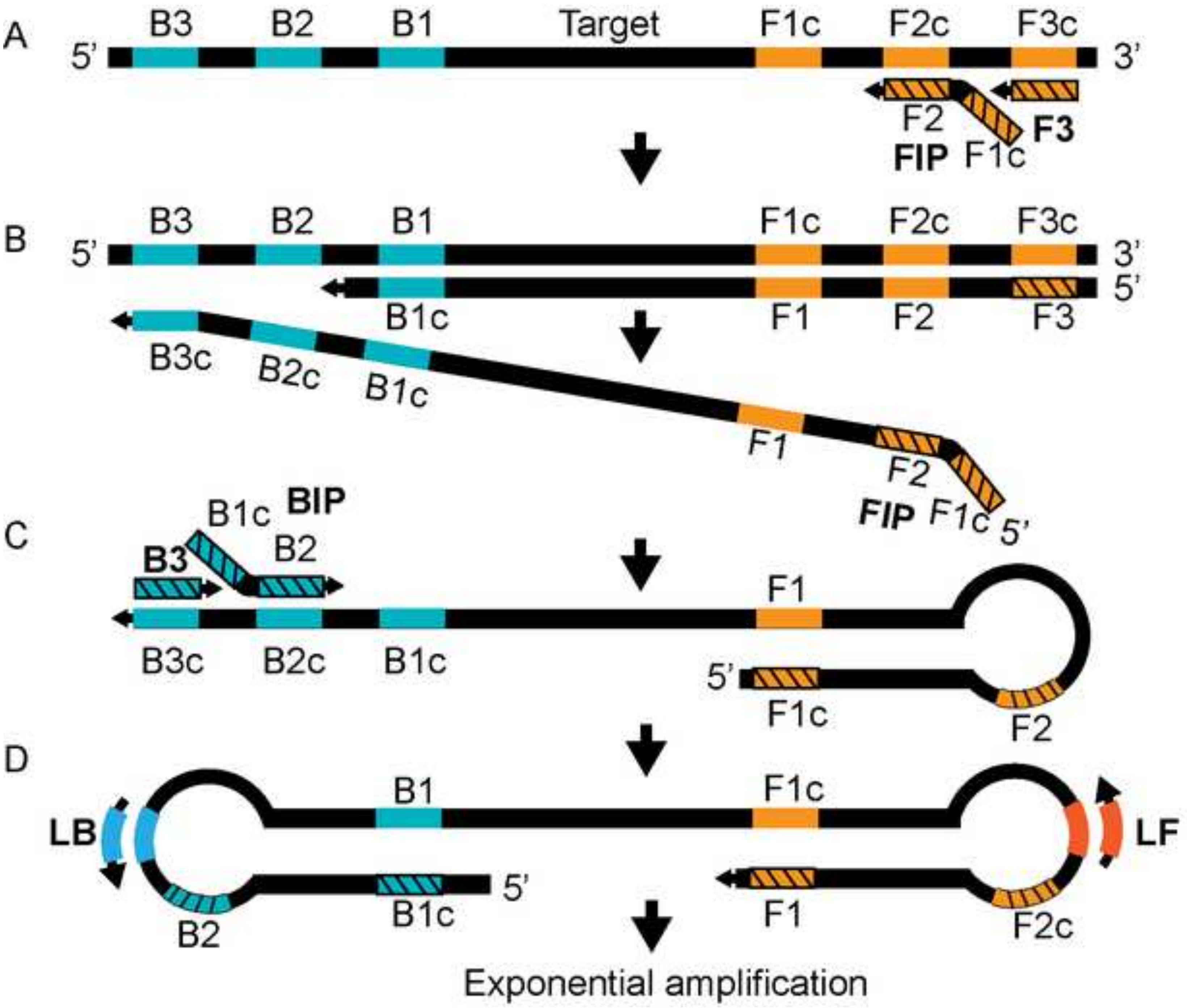

Figure 5

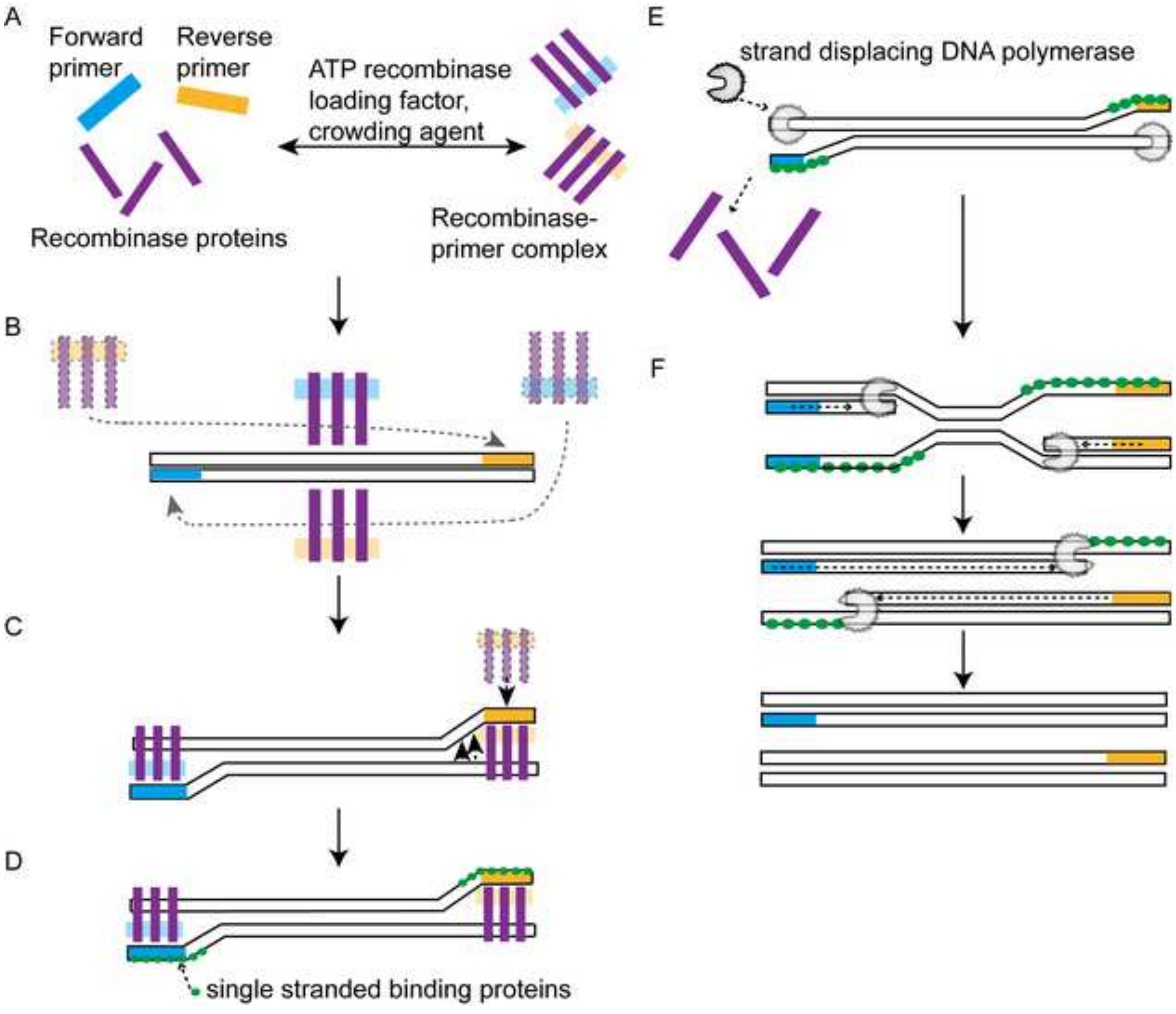
A
Forward primer
Reverse primer
ATP recombinase loading factor, crowding agent
Recombinase proteins
Recombinase-primer complex
B
C
D
single stranded binding proteins
E
strand displacing DNA polymerase
F

**Figure 1.** Sturcture and genome organization of MPXV. **(**A) Structure of MPXV. Redrawn from Ref. (Frey and Belshe, 2004). (B) Schematic representation of MPXV genome of the strain Zaire-96-1-16 isolated during the 1996 outbreak in Zaire. Redrawn from Ref. (Saxena et al., 2022)

**Figure 2.** Multi-country MPXV outbreak. Numbers indicate the total number of cases in each country during Janurary, 2022-June, 2022. Redrawn from Ref. (WHO, 2022b), (WHO, 2022e).

**Figure 3.** Schematic of the image-based MPXV detection workflow. Redrawn from Ref. (Ahsan et al., 2022b).

**Figure 4.** Reaction mechanism of LAMP. Redrawn from Ref. (Becherer et al., 2020).

**Figure 5.** Reaction mechanism of RPA. Recombinase complexation with primer (A), scanning of homologous sequences by recombinase-primer complex (B). Strand displacement by recombinase and primer insertion (C) and binding of single strand binding proteins to stabilize the primer binding (D). Recombinase disassembly and binding of strand displacing DNA polymerase (E). Elongation reaction (F). Redrawn from Ref. (Lobato and O’Sullivan, 2018).

**Table 1 Summary of MPXV detection methods**

| Sr. No | Assay name | Target gene | Primers' sequence | Probes' sequence | Detection limit | Real-sample analysis | Ref. |
|---|---|---|---|---|---|---|---|
| 1 | HA-PCR | HA gene | **Forward**-5′-CTGATAATGTAGAAG AC -3′ **Reverse**-5′-TTGTATTTACGTGGGTG-3′ | NA | Not reported | Yes | (Ropp et al., 1995) |
| **2** | ATI-PRC | ATI-gene | **Forward**-5′-AATACAAGGAGGATCT-3′ **Reverse**-5′-CTTAACTTTTTCTTTTTCTTTCTC-3′ | NA | Not reported | Yes | (Meyer et al., 1997) |
| **3** | MPXV PCR assay | ATI-gene | **Forward** -5′-GAGAGAATCTCTTGATAT-3′ **Reverse**-5′-ATTCTAGATTGTAATC-3′ | NA | Not reported | Yes | (Neubauer et al., 1998) |
| **4** | Real-time PCR | B6R | **Forward** 5′-ATTGGTCATTATTTTTGTCACAGGAACA-3′ | 5′-MGB/DarkQuencher-AGAGATTAGAAATA-3′-FAM | ~10 viral copies (2 fg) | Yes | (Li et al., 2006) |
| | | | **Reverse**-5′-AATGGCGTTGACAATTATGGGTG-3′ | | | | |
| **5** | Real-time PCR | TNF | **Forward**-5′-CACACCGTCTCTTCCACAGA -3′ **Reverse**-5′-GATACAGGTTAATTTCCACATCG -3′ | 5′-FAM AACCCGTCGTAACCAGCAATACATTT-3′-BHQ1 | ~8.2 genome copies (1.7 fg) | Yes | (Li et al., 2010b) |
| **6** | Real-time PCR | TNF | **Forward** -5 ′ -TGTCTACCTGGATACAGAAAGCAA-3′ **Reverse**-5′-GGCATCTCCGTTTAATACATTGAT -3′ | 5′-FAM-CCCATATATGCTAAATGTACCGGTACCGGA-3′-BHQ1 | ~40.4 copies (9.46 fg) | Yes | (Li et al., 2010b) |
| **7** | Real-time PCR | F3L | **Forward**-5′-CTCATTGATTTTTCGCGGGAT A-3′ **Reverse**-5′-GACGATACTCCTCCTCGTTGGT-3′ | 5′-6FAM-CATCAGAATCTGTAGGCCGT-MGBNFQ-3′ | 11–55 fg (50–250 copies) | Yes | (Kulesh et al., 2004b) |
| **8** | Real-time PCR | N3R | **Forward-** 5 ′ -AACAACCGT CCTACA ATTAAA CAACA-3′ | 5′-6FAM-TAT AAC GGC GAA GAA TAT ACT-MGBNFQ-3′ | 11–55 fg (50–250 copies) | Rodents | (Kulesh et al., 2004b) |
| | | | **Reverse-** 5′-CGCTATCGAACCATTTTTGTAGTCT-3′ | | | | |
| **9** | Real-time PCR | B7R | Forward -5′-ACGTGTTAAACAATGGGTGATG-3′<br>Reverse- 5′-AACATTTCCATGAATCGTAGTCC-3′ | 5′-TAMRA-TGAATGAATGCGATACTGTATGTGTGGG-3′-BHQ2 | 50 copies per reaction | Yes | (Shchelkunov et al., 2011) |
| **10** | C-LAMP | D14L | **FIP-C-5'-**TGGGAGCATTGTAACTTATAGTTGCCCTCCTGAACACATGACA-3'<br>**F3-**C-5'-TGGGTGGATTGGACCATT-3'<br>**BIP-C-**5'-ATCCTCGTATCCGTTATGTCTTCCCACCTATTTGCGAATCTGTT-3'<br>**B3-C-**5'-ATGGTATGGAATCCTGAGG-3'<br>**LOOP-F-C-**5'-GATATTCGTTGATTGGTAACTCTGG-3' | N/A | $10^{2.4}$ copies per reaction | Yes | (Iizuka et al., 2009) |
| | | | **LOOP-C-C-**5’-GTTGGATATAGATGGAGGTGATTGG-3’ | | | | |
| **11** | C-LAMP | ATI | **FIP-W-**5’-CCGTTACCGTTTTTACAATCGTTAATCAATGCTGATATGGAAAAGAGA-3’<br>**F3-W-**5’-TACAGTTGAACGACTGCG-3’<br>**BIP-W-**5’-ATAGGCTAAAGACTAGAATCAGGGATTCTGATTCATCCTTTGAGAAG-3’<br>**B3-W-**5’-AGTTCAGTTTTATATGCCGAAT-3’<br>**LOOP-F-W-**5’-GATGTCTATCAAGATCCATGATTCT-3’<br>**LOOP-C-W-**5’-TCTTGAACGATCGCTAGAGA-3’ | N/A | $10^3$ copies per reaction | Yes | (Iizuka et al., 2009) |
| **12** | RPA | TNF (G2R) | **Forward-** 5’-AATAAACGGAAGAGATATAGCACCACATGCAC-3’ **Reverse-**5’-GTGAGATGTAAAGGTATCCGAACCACACG-3’ | 5’-ACAGAAGCCGTAATCTATGTTGTCTATCGQTFCCTCCGGGAACTTA-3’ | 16 DNA molecules/µl | Yes | (Davi et al., 2019) |